\documentclass[11pt]{amsart}
\usepackage[foot]{amsaddr}

\usepackage[a4paper, margin=2cm]{geometry}

\usepackage[super,compress]{cite}
\usepackage{graphicx}

\usepackage[dvipsnames]{xcolor}
\usepackage{graphicx}
\usepackage{amssymb}
\usepackage{epstopdf}
\usepackage{float}
\usepackage{subcaption}

\usepackage{listings}
\usepackage{amsmath}
\usepackage{verbatim}
\usepackage{url}
\usepackage{tikz}
\usetikzlibrary{arrows,arrows.meta,positioning}

\usepackage{array}
\newcolumntype{L}[1]{>{\raggedright\let\newline\\\arraybackslash\hspace{0pt}}m{#1}}
\newcolumntype{C}[1]{>{\centering\let\newline\\\arraybackslash\hspace{0pt}}m{#1}}
\newcolumntype{R}[1]{>{\raggedleft\let\newline\\\arraybackslash\hspace{0pt}}m{#1}}

\usepackage{todonotes}

\lstdefinelanguage{julia}
{
  keywordsprefix=\@,
  morekeywords={
    exit,whos,edit,load,is,isa,isequal,typeof,tuple,ntuple,uid,hash,finalizer,convert,promote,subtype,typemin,typemax,realmin,realmax,sizeof,eps,promote_type,method_exists,applicable,invoke,dlopen,dlsym,system,error,throw,assert,new,Inf,Nan,pi,im,begin,while,for,in,return,break,continue,macro,quote,let,if,elseif,else,try,catch,end,bitstype,ccall,do,using,module,import,export,importall,baremodule,immutable,local,global,const,Bool,Int,Int8,Int16,Int32,Int64,Uint,Uint8,Uint16,Uint32,Uint64,Float32,Float64,Complex64,Complex128,Any,Nothing,None,function,type,typealias,abstract,struct, mutable
  },
  sensitive=true,
  morecomment=[l]{\#},
  morestring=[b]',
  morestring=[b]" 
}
\begin{document}

\title{Integration with an Adaptive Harmonic Mean Algorithm}


\author{Allen Caldwell$^1$, Philipp Eller$^2$, Vasyl Hafych$^1$, Rafael Schick$^{1,2}$, Oliver Schulz$^1$, Marco  Szalay$^{1,3}$}
\address{1: Max Planck Institute for Physics, D-80805 Munich, Germany\\
2: Technical University of Munich, D-80333 Munich, Germany\\
3: now at Google LLC., Santa Barbara, CA, 93101, USA}

\maketitle

\begin{abstract}
Numerically estimating the integral of functions in high dimensional spaces is a non-trivial task. A oft-encountered example is the calculation of the marginal likelihood in Bayesian inference, in a context where a sampling algorithm such as a Markov Chain Monte Carlo provides samples of the function. We present an Adaptive Harmonic Mean Integration (AHMI) algorithm. Given samples drawn according to a probability distribution proportional to the function, the algorithm will estimate the integral of the function and the uncertainty of the estimate by applying a harmonic mean estimator to adaptively chosen regions of the parameter space. We describe the algorithm and its mathematical properties, and report the results using it on multiple test cases.

\end{abstract}

\section{Introduction}

Sampling algorithms, such as Markov Chain Monte Carlo (MCMC)~\cite{ref:MCMC}, are often used to generate samples distributed according to non-trivial densities in high dimensional spaces. Many algorithms have been developed that allow generating samples $\{\Lambda\}$ from an unnormalized target density $f(\lambda)$:
$$ \Lambda \sim f(\lambda)\;\;\;\;\;\;\;\;\;\;\;\;\;\;\;\; f(\lambda)\geq 0 \;\;\; .$$
In many applications, it is desirable or even necessary to be able to normalize the target density; i.e., to calculate
\begin{equation}
\label{eq:integral}
 I  \equiv  \int_{\Omega}  f(\lambda) d\lambda 
\end{equation}
where $\Omega$ is the support of $f$.  This integral can be computationally very costly or impossible to compute using standard techniques; e.g., if the volume of the support where the target $f$ is non-negligible occupies a very small region of the full support.  

An important area where such integration is necessary is Bayesian inference~\cite{ref:Jeffreys,ref:Jaynes}.  Bayes' formula, for a given model $M$, is
\begin{equation}
P(\lambda | {\rm Data},M) = \frac{ P({\rm Data}| \lambda,M) P_0(\lambda|M) }{P({\rm Data}|M)}
\end{equation}
where here $\lambda$ denotes the parameters of the model and the data are used to update probabilities for possible values of $\lambda$ from prior probabilities $P_0(\lambda|M)$ to posterior probabilities $P(\lambda | {\rm Data},M)$. The denominator is usually expanded using the Law of Total Probability and written in the form 
\begin{equation}
Z=P({\rm Data}|M)=\int P({\rm Data}|\lambda,M) P_0(\lambda|M) d\lambda.
\end{equation}
Z goes by the name `evidence', or `marginal likelihood', and is an example of the type of integral that we want to be able to calculate (here the data are fixed and $f(\lambda) = P({\rm Data}|\lambda,M) P_0(\lambda|M) $). An example use of $Z$ is the calculation of a Bayes factor to compare two models:
$$
{\rm BF} \equiv \frac{P({\rm Data}| M_A)}{P({\rm Data}| M_{B})} = \frac{Z_A}{Z_{B}} \; .
$$

We are specifically interested in providing an algorithm applicable in a setting where samples are available from the target density $f(\lambda)$ but with possibly no further recourse to generating more samples. For this purpose, we investigate the use of a modified harmonic mean estimator (HME)~\cite{ref:HME}.  We introduce the use of a reduced integration region to improve the HME performance.  After a description of the technique, we report on numerical investigations using Metropolis-Hastings MCMC samples.  
Our work has in common with ~\cite{ref:Weinberg} the use of the ratio of samples found in the limited support region to the number found in the full support, but we use a substantially different integral estimation technique.

\section{Reduced Volume HME}
We are interested in estimating the integral $I$ from Eq.~\ref{eq:integral}.  We start by defining the integral
\begin{equation}
 I_\Delta  \equiv  \int_{\Delta}  f(\lambda) d\lambda 
\end{equation}
with $\Delta \subset \Omega$ a finite integration region, and the ratio
\begin{equation}
\label{eq:scheme}
r \equiv  \frac{I_\Delta}{I}.
\end{equation}
Given our assumption that the sampling algorithm has successfully sampled from $f(\lambda)$, we use the following as an 
estimator to our ratio
\begin{equation}
\label{eq:scheme2}
\hat{r} = \frac{N_{\Delta}}{N_{\Omega}},
\end{equation}
which is the fraction of the total number of samples that fall within $\Delta \subset \Omega$ .
Defining the normalized density over $\Delta$ 
\begin{equation}
 \tilde{f}_{\Delta}(\lambda) = \frac{f(\lambda)}{I_\Delta} \;\;\;\;\;\;\;\;\; \lambda \in \Delta ,
\end{equation}
allows us to perform a harmonic mean calculation as follows:
\begin{eqnarray}
E\left[ \frac{1}{f(\lambda)}\right]_{\tilde{f}_{\Delta}(\lambda)} = \int_{\Delta} \frac{1}{f(\lambda)} \cdot \tilde{f}_{\Delta}(\lambda) d\lambda = \int_{\Delta} \frac{1}{f(\lambda)} \cdot \frac{f(\lambda)}{I_{\Delta}} d\lambda = \frac{V_{\Delta}}{I_{\Delta}} 
\end{eqnarray}
where $V_{\Delta}$ is the volume of the region defined by $\Delta$. An estimator for this expectation value is the harmonic mean
\begin{equation}
\hat{X}=\frac{1}{N_{\Delta}}\sum_{\lambda_i \in \Delta} \frac{1}{f(\lambda_i)}.
\end{equation}
The HME for the reduced volume integral then follows as
\begin{equation}
\label{eq:red}
\hat{I}_{\Delta} = \frac{V_{\Delta}}{\hat{X}} =  \frac{N_{\Delta} V_{\Delta}}{ \sum_{\lambda_i \in \Delta} \frac{1}{f(\lambda_i)}} \; .
\end{equation}
This calculation is performed directly from the values of the target density $f(\lambda_i)$ given by the sampling algorithm, and does not require any extra sampling.
An estimator for the integral over the full space $\Omega$ can then be written down as
\begin{equation}
\label{eq:ourHME}
\hat{I} = \frac{\hat{I}_{\Delta}}{\hat{r}} =\frac{N_{\Omega} V_{\Delta}}{ \sum_{\lambda_i \in \Delta} \frac{1}{f(\lambda_i)}}\; .
\end{equation}
The task of estimating our integral therefore reduces to choosing one or several sub-spaces $\Delta$---typically small regions around local modes of $f(\lambda)$. 
The full space $\Omega$ over which the integration ought to be performed can be large or even infinite, while this does not affect the outcome of our integral estimate. We discuss the bias and uncertainty of this estimator in the following subsection.

In general, MCMC samples come with weights (e.g. repeated samples, with the weight being the number of repetitions). We therefore rewrite Eq.~\ref{eq:ourHME} as
\begin{equation}
\label{eq:ourHME2}
\hat{I} = \frac{W_{\Omega} V_{\Delta}}{ \sum_{\lambda_i \in \Delta} \frac{w_i}{f(\lambda_i)}}
\end{equation}
with $w_i$ the weights assigned to the samples at parameter values $\lambda_i$ and $W_{\Omega} = \sum_i w_i$ the sum of all weights.
The use of weights also allows this technique to be applied to samples obtained from, for example, importance sampling.

\subsection{Illustration of the Technique}
\label{sec:illustration}

\begin{figure}
\centering
\includegraphics[width = \textwidth]{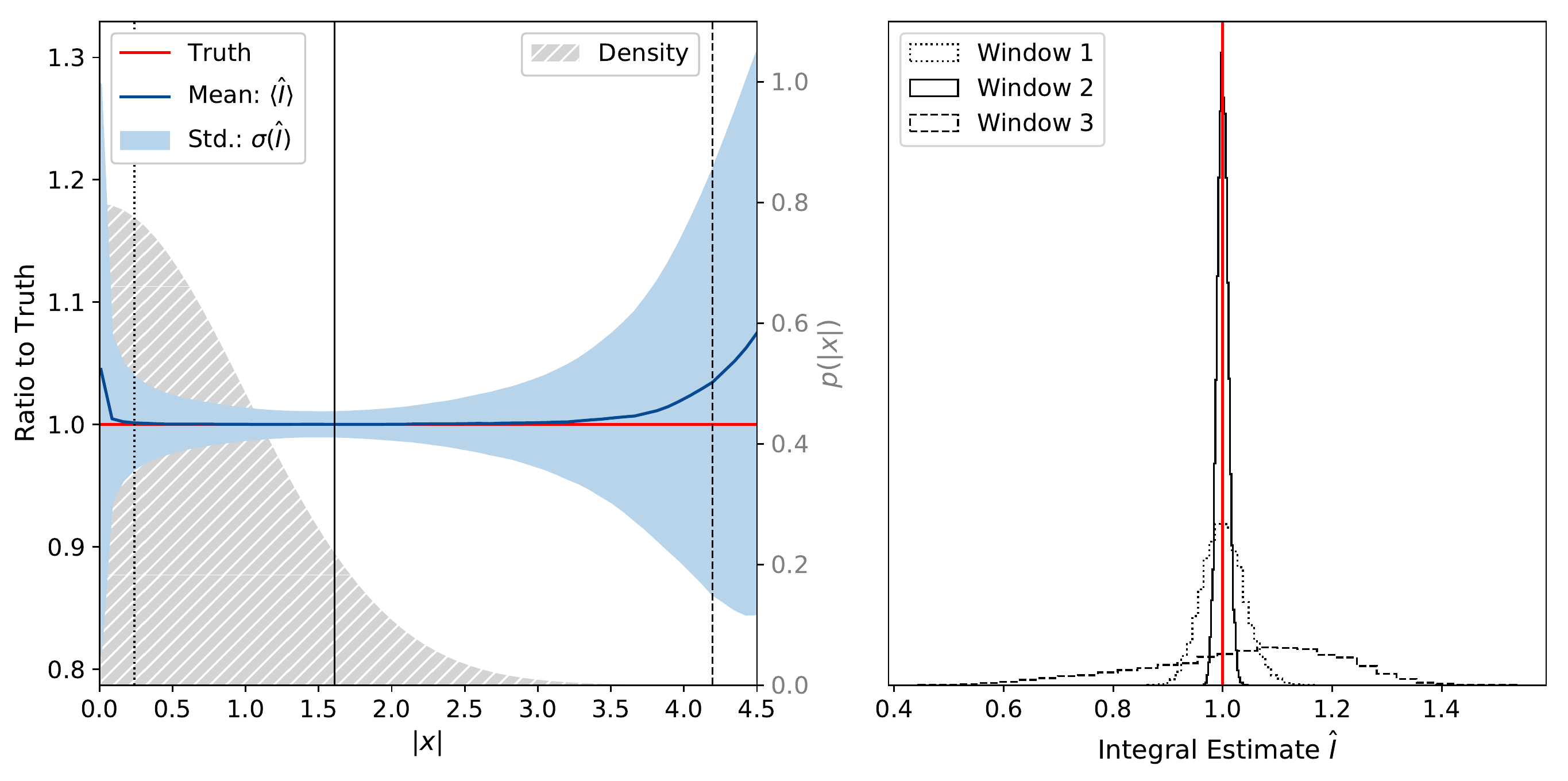}
\caption{Demonstration of the reduced harmonic mean technique with the unit normal distribution.
The left panel shows in blue the mean and the $\pm1\sigma$ region (from $1.5\cdot10^4$ repeated trials) of the integral estimates as a function of the window extent. The true integral value is 1.0, indicated in red. The gray shaded distribution shows the unit normal pdf (right y-axis scale). For the three window extents indicated by the black, vertical lines (Window 1: $|x| < 0.24$, Window 2: $|x| < 1.61$, Window 3: $|x| < 4.20$) the full distribution of integral estimates from the $1.5\cdot10^4$ trials are plotted on the right side as histograms.}
\label{fig:demo}
\end{figure}

To illustrate our technique for applying harmonic mean integration, we consider the unit normal distribution $p(x) = \mathcal{N}(x|\mu=0, \sigma=1) $.  A fixed number of samples ($3\cdot10^3$) was generated from directly sampling the unit normal distribution, and Eq.~\ref{eq:ourHME} was used to calculate the integral for different sub-regions $\Delta$.  These regions are defined as windows of $x$ centered on $0$ and varied in width from $0.02$ up to $9$.  This was repeated $1.5\cdot10^4$ times, and the mean and standard deviation were evaluated. Figure~\ref{fig:demo} shows the results of the integration as a function of window size.  As is seen in the figure,  harmonic mean integration applied to a finite region gives an accurate value for the integral over a wide range of sampling windows.  The variation in the integral results is largest for small windows due to the small number of samples used, and for large windows due to the divergence of the harmonic mean estimator. We discuss the biases in the next section.

\subsection{Bias and Uncertainty of the Estimator}

\begin{figure}
\centering
\includegraphics[width = \textwidth]{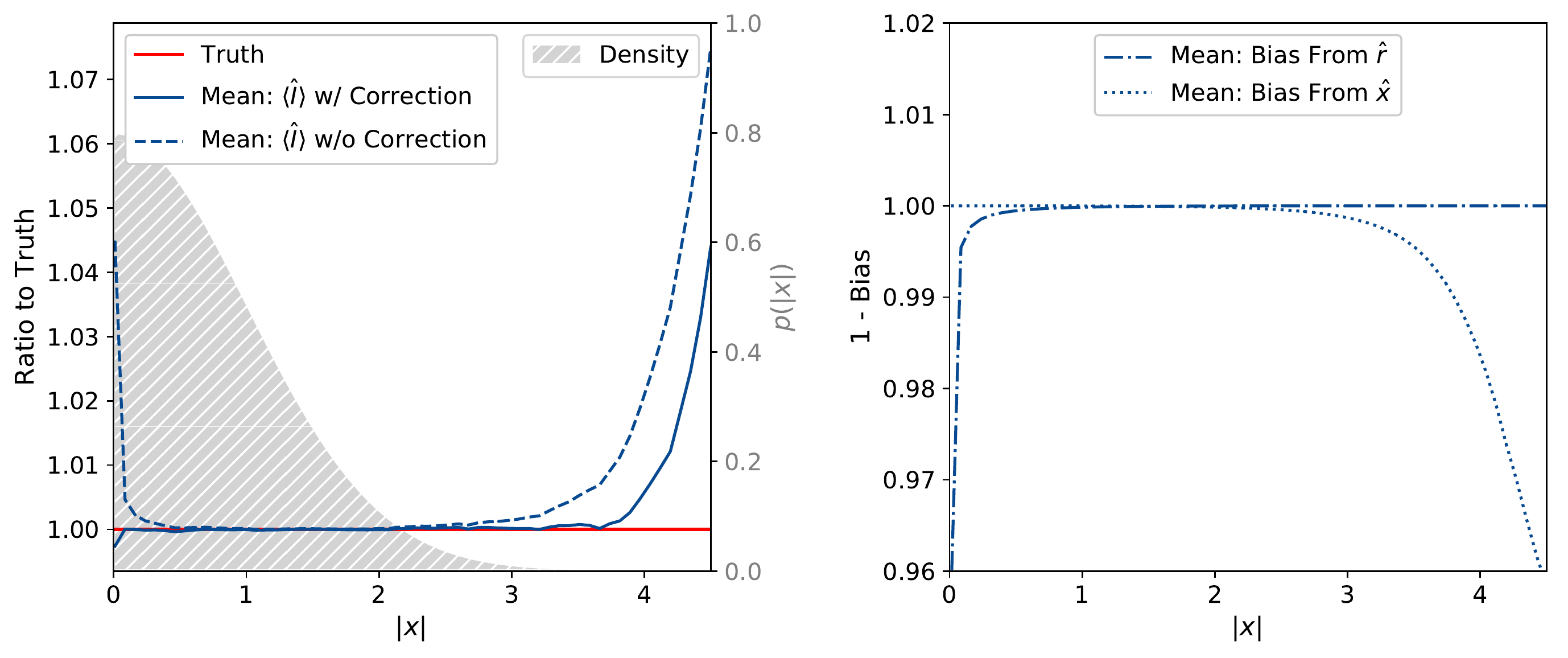}
\caption{Left: The average uncorrected (Eq.~\ref{eq:ourHME}) and corrected ($b\cdot\hat{I}$) integrals as a function of the extent of the window used to accept samples. The results are from averaging $1.5\cdot10^4$ integration results, where each integration test was performed from $3\cdot10^3$ samples in the full function range.
Right: Individual contributions to the bias correction from the Binomial ($\hat{r}$) and the $1/f$ terms.
}   
\label{fig:correction}
\end{figure}

We can estimate the bias and uncertainty on $\hat{I}$ given in Eq.~\ref{eq:ourHME} by separately analyzing the behavior of $\hat{r}$ and $\hat{X}$. As described below, we choose regions $\Delta$ for which the range of target density values is moderate. For \textit{i.i.d.} sampling, this would imply, via the Central Limit Theorem, that $\hat{X}$ follows a Normal distribution. Assuming we can approximate the distribution of $\hat{X}$ with a Normal distribution, we have
$$
P(\hat{X})\approx  \mathcal{N}\left(\hat{\mu}_X=\hat{X},\hat{\sigma}^2_X = \frac{\sum_{\lambda_i\in \Delta}(\frac{1}{\lambda_i} - \hat{X})^2}{N_{\Delta}(N_\Delta-1)}\right)
$$
where $P(\hat{X})$ is the probability distribution for $\hat{X}$ with mean $\mu$ and variance $\sigma^2$ estimated from the observed samples.  
Since $\hat{X}$ appears in the denominator in Eq.~\ref{eq:ourHME}, this produces a bias in our integral of size $\hat{\sigma}^2_X/\hat{\mu}^2_X$.  The fractional uncertainty in our integral estimator is $\hat{\sigma}_X/\hat{\mu}_X$.

The estimator $\hat{r}$ will also typically follow approximately a Normal distribution with parameters that can be estimated from \textit{i.i.d.} sampling and Binomial statistics as
$$
P(\hat{r})\approx  \mathcal{N}\left(\hat{\mu}_r=\hat{r},\hat{\sigma}^2_r = \frac{\hat{r}(1-\hat{r})}{N_{\Omega}}\right)
$$
Since $\hat{r}$ also appears in the denominator, it will also produce a bias in our integral of size $\hat{\sigma}^2_r/\hat{\mu}^2_r$.  The fractional uncertainty in our integral estimator from $\hat{r}$ is, in the approximation of \textit{i.i.d.} sampling and a Normal distribution, $\hat{\sigma}_r/\hat{\mu}_r$.

We can therefore write down an explicit correction factor 
\begin{equation}
    b = (1 - \frac{\hat{\sigma}^2_X}{\hat{\mu}^2_X} - \frac{\hat{\sigma}^2_r}{\hat{\mu}^2_r})
\end{equation}
that we apply to the integral estimate $\hat{I}$.

The correction is illustrated in Fig.~\ref{fig:correction} using the same numerical experiments discussed in section~\ref{sec:illustration}.
The resulting uncorrected and corrected average integral values are displayed. Focusing on small window sizes, it can be seen that the term from $\hat{r}$ dominates, as the binomial uncertainty is largest for small numbers of samples. With the bias correction applied, already for as few as $\approx 20$ samples inside the integration volume (corresponding to roughly $|x| < 0.01$), correct results are produced, while without the correction factor applied much large windows, starting at around $|x| < 0.5$, are necessary.

Towards larger windows the bias produced from $\hat{X}$ become dominant, as the range of values of $f$ of the contained samples grows. The bias correction successfully mitigates this effect as illustrated.

Once the window size exceeds the space where samples are present, the integral starts diverging. For our example using $3\cdot10^3$ samples, we expect those to cover a region up to only $|x|\approx3.58$, which explains well the observed trend. 


Many samplers such as MCMC algorithms generate strong correlations amongst samples and using the binomial uncertainty discussed here can be inaccurate. We therefore also numerically evaluate the uncertainty as described in detail below.  The integration regions are chosen such that the bias correction can be neglected.

\subsection{Relation to other Techniques for Evidence Calculation}
A variety of techniques to calculate the marginal likelihood in Bayesian calculations have been successfully developed.  A summary can be found in~\cite{ref:evidence}, where a number of MCMC related techniques are reviewed, including Laplace's method~\cite{ref:Laplace}, harmonic mean estimation~\cite{ref:HME}, Chib's method~\cite{ref:Chib}, annealed importance sampling techniques~\cite{ref:Neal,ref:Robert}, nested sampling~\cite{ref:Skilling} and thermodynamic integration methods~\cite{ref:thermo,ref:Friel}. Only the HME and Laplace techniques allow the direct estimation of the evidence from available samples, and the Laplace technique makes the unwanted assumption that the target density is a multi-variate Gaussian.

In the Bayesian literature~\cite{ref:HME}, the HME for the evidence, $Z$, is formulated as
\begin{eqnarray}
\label{eq:BayesHME}
Z &=&\frac{1}{E\left[\frac{1}{P(\rm Data|\lambda,M)}\right]_{P(\lambda|{\rm Data},M)}} \\
&=&\frac{1}{\int \frac{P(\lambda|{\rm Data},M)}{P(\rm Data|\lambda,M)} d\lambda} \; .
\end{eqnarray}
 The difference with our formulation is that the prior $P_0(\lambda|M)$ is not included in the expectation value, and the integration volume does not appear.  In order to perform this HME calculation, it is necessary to know separately the value of the likelihood and the prior at the sampling points.  This method has been strongly criticized (even called ``worst Monte Carlo Method ever''~\cite{ref:Neal2}), since the evaluation of the denominator in Eq.~\ref{eq:BayesHME} can have very large variance.  Reducing the variance by limiting the integration region is not as straightforward as in our formulation since it requires an integral of the prior function over the reduced support.

\section{An Adaptive Harmonic Mean Integration Algorithm}
\label{sec:adaptiveharmonicmean}

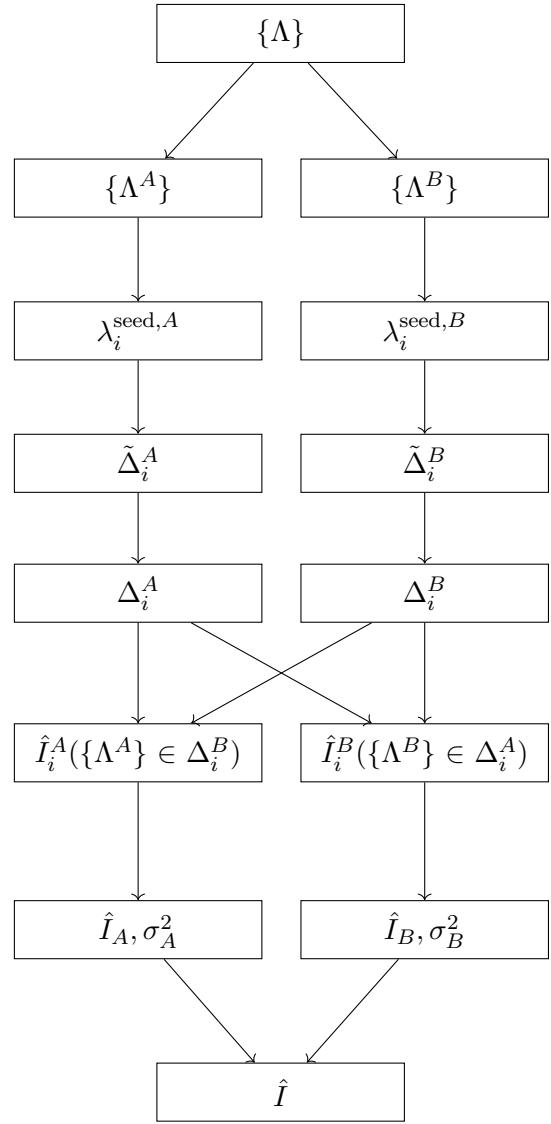
\begin{figure}[b]
\centering

\tikzstyle{block} = [rectangle, draw, fill=white, text width=3cm, text centered, minimum height=2em]
\tikzstyle{plain} = [draw=none,fill=none, text width=0.4\textwidth]

\begin{tikzpicture}[] 

\node [plain] (t0) {(1) Start with samples $\{\Lambda\}$ from density $f(\lambda)$ and apply whitening transformation};
\node [block, right = 2.375cm of t0] (b0) {$\{\Lambda\}$};

\node [plain, below = 1.5em of t0] (t2) {(2) Split into two mutually exclusive, uncorrelated and equally sized sub-sets $A$ \& $B$};
\node [block, right = 0.5cm of t2] (b2a) {$\{\Lambda^A\}$};
\node [block, right = 0.5cm of b2a] (b2b) {$\{\Lambda^B\}$};

\node [plain, below = 1.5em of t2] (t3) {(3) Generate seed points $\lambda^{\textrm{seed}}_i$ (via space partitioning tree)};
\node [block, right = 0.5cm of t3] (b3a) {$\lambda^{\textrm{seed},A}_i$};
\node [block, right = 0.5cm of b3a] (b3b) {$\lambda^{\textrm{seed},B}_i$};

\node [plain, below = 1.5em of t3] (t5) {(4) Create a small hyper-cube $\tilde{\Delta}_i$ around each seed point $\lambda^{\textrm{seed}}_i$};
\node [block, right = 0.5cm of t5] (b5a) {$\tilde{\Delta}_i^A$};
\node [block, right = 0.5cm of b5a] (b5b) {$\tilde{\Delta}_i^B$};

\node [plain, below = 1.5em of t5] (t7) {(5) Adjust the faces of the hyper-cubes, resulting in hyper-rectangles};
\node [block, right = 0.5cm of t7] (b7a) {$\Delta_i^A$};
\node [block, right = 0.5cm of b7a] (b7b) {$\Delta_i^B$};

\node [plain, below = 1.5em of t7] (t8) {(6) Use the resulting hyper-rectangles $\Delta^B_i$ created with sample $\{\Lambda^B\}$ for the harmonic mean estimates $\hat{I}^A_i$ on $\{\Lambda^A\}$ and vice-versa};
\node [block, right = 0.5cm of t8] (b8a) {$\hat{I}_i^A(\{\Lambda^A\} \in \Delta_i^B)$};
\node [block, right = 0.5cm of b8a] (b8b) {$\hat{I}_i^B(\{\Lambda^B\} \in \Delta_i^A)$};

\node [plain, below = 1.5em of t8] (t9) {(7) Use individual estimates $\hat{I}_i$ to compute combined estimate and variance  for $A$ \& $B$};
\node [block, right = 0.5cm of t9] (b9a) {$\hat{I}_A, \sigma^2_A$};
\node [block, right = 0.5cm of b9a] (b9b) {$\hat{I}_B, \sigma^2_B$};

\node [plain, below = 1.5em of t9] (t10) {(8) Compute final estimate by combining the independent estimates $\hat{I}_A$ and $\hat{I}_B$ weighted by their variance.};
\node [block, right = 2.375cm of t10] (b10) {$\hat{I}$};

\draw [->] (b0) -- (b2a);
\draw [->] (b0) -- (b2b);
\draw [->] (b2a) -- (b3a);
\draw [->] (b2b) -- (b3b);
\draw [->] (b3a) -- (b5a);
\draw [->] (b3b) -- (b5b);
\draw [->] (b5a) -- (b7a);
\draw [->] (b5b) -- (b7b);
\draw [->] (b7a) -- (b8a);
\draw [->] (b7a) -- (b8b);
\draw [->] (b7b) -- (b8a);
\draw [->] (b7b) -- (b8b);
\draw [->] (b8a) -- (b9a);
\draw [->] (b8b) -- (b9b);
\draw [->] (b9a) -- (b10);
\draw [->] (b9b) -- (b10);

\end{tikzpicture}

\caption{Overview of the different steps in the AHMI algorithm, including the procedure of finding subvolumes $\Delta_i$ and computing integral estimates $\hat{I}_i$.}
\label{fig:flowchart}
\end{figure}

Adaptive Harmonic Mean Integration (AHMI) uses the HME on multiple subregions $\Delta_i$ to estimate the integral of $f(\lambda)$ over its full support $\Omega$. In this section we present our example algorithm in detail and will show benchmark tests on several distributions in Sec.~\ref{sec:benchmark}. As discussed previously, defining a set of suitable regions is crucial in obtaining a robust and unbiased estimate of the integral of $f(\lambda)$. In particular, to avoid biasing the result, it is essential not to use the same elements of the sample set $\{\Lambda\}$ for both the definition of $\Delta_i$ and estimates of the integral $\hat{I}_i$.

The general flow of the AHMI algorithm, including the procedure of defining the $\Delta_i$, is summarized in Fig.~\ref{fig:flowchart}. The various involved steps are discussed in more technical detail in the following subsections.

\subsection{Samples, Preprocessing and Splitting}

We start with a given set of samples $\{\Lambda\}$ that we assume are drawn according to the probability distribution proportional to our function $f$, obtained for example from MCMC sampling.

In order to de-correlate the sample space we apply a whitening transformation.
In general, a whitening transformation maps a set of random variables with a known non-singular covariance matrix to a new set of variables with a covariance matrix equal to $\mathbb{I}$.
A Cholesky Decomposition is used to whiten the samples, and the AHMI estimator for the integral becomes
\begin{equation}
\hat{I}  =\frac{W_{\Omega} V'_{\Delta}}{ \det R \cdot \sum_{\lambda'_i \in \Delta} \frac{w_i }{f(\lambda'_i)}}
\end{equation}
where $\det R$ is the determinant of the whitening matrix and the primed symbols represent the quantities in the transformed space.
In the following we drop this explicit addition of prime symbols and work in the whitened space (unless otherwise stated).

The full set of samples is then divided into two equally sized subsets $A$ and $B$. It is important that the resulting subsets are independent; one way to achieve that is to use MCMC samples from separate chains.

\subsection{Hyper-rectangle Generation}

\begin{figure}[t!]
\centering
    \begin{subfigure}[t]{0.45\textwidth}
    \includegraphics[width=\linewidth]{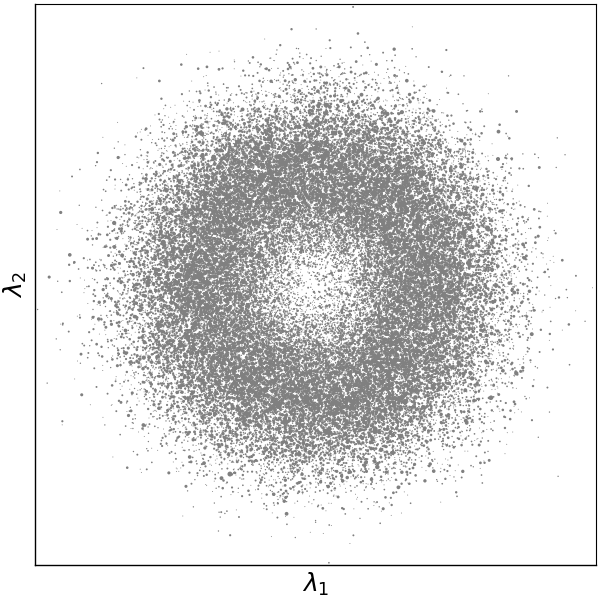}
    \caption{MCMC samples after whitening transformation. The size of each point is proportional to its weight.}
    \label{fig:samples}
    \end{subfigure}
    \hspace{1em}
    \begin{subfigure}[t]{0.45\textwidth}
    \centering
    \includegraphics[width=\linewidth]{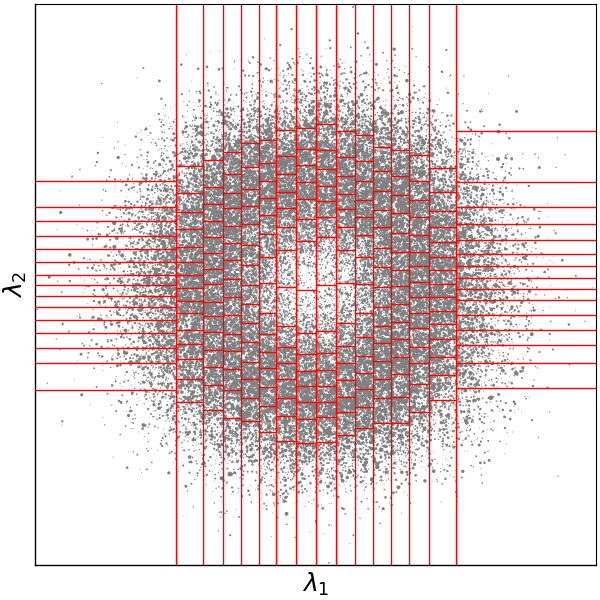}
    \caption{Two dimensional space-partitioning tree. All regions contain an equal number of unique samples.}
    \label{fig:spacepartitioningtree}
    \end{subfigure}
    
    \vspace{1cm}
    
    \begin{subfigure}[t]{0.45\textwidth}
    \centering
    \includegraphics[width=\linewidth]{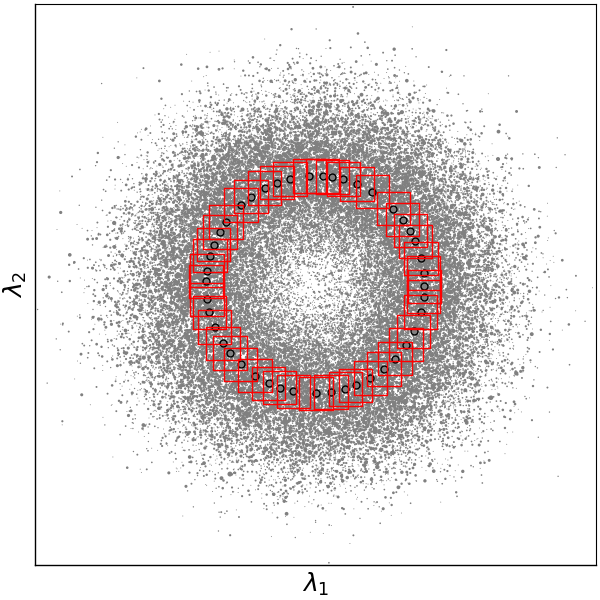}
    \caption{Initial hypercubes $\tilde{\Delta}$ around seed points.}
    \label{fig:seed}
    \end{subfigure}
    \hspace{1em}
    \begin{subfigure}[t]{0.45\textwidth}
    \centering
    \includegraphics[width=\linewidth]{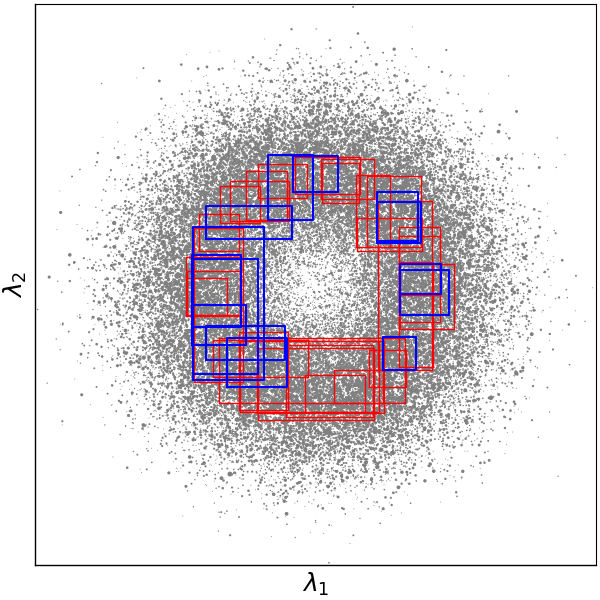}
    \caption{Hyper-rectangles after adjustments. Only the red hyper-rectangles are used in the final integral estimate.
    }
    \label{fig:visualizationhmi}
    \end{subfigure}
\caption{Process of finding integration regions in a two dimensional example for the Gaussian shells test function.}
\label{fig:subvolumes}
\end{figure}

We illustrate the hyper-rectangle generation steps in more detail using a two-dimensional Gaussian shell example with distribution
\begin{equation}
\label{2dshell}
f(\lambda| \vec{c},r,\omega)=\frac{1}{\sqrt{2\pi\omega^2}}\exp\left(-\frac{(\vert \lambda-\vec{c}\vert-r)^2}{2\omega^2}\right)\; .
\end{equation}

In our examples, we use the following settings: radius $r=5$, width $\omega=2$ and $\vec{c}=\vec{0}$.  The integration region extends from $[-25,25]$ in each dimension.  Samples from this distribution are shown in Fig.~\ref{fig:samples}.


The algorithm starts by creating seed points around which to construct the integration regions~$\Delta_i$. These points should lie in areas of high density and should result in broadly distributed starting points. 
In order to limit computation time, a simple space-partitioning tree is used to divide the whitened space into subsets of non-overlapping regions with an equal number of samples inside them. An example of such a space-partitioning tree is shown in Fig.~\ref{fig:spacepartitioningtree}.
For each partition $i$, the sample with the largest function value contained in that partition is defined as seed point $\lambda^{seed}_i$.

The following steps produce hyper-rectangle-shaped regions suitable for AHMI.
In order to limit this variance and ensure numerical stability, the ratio between the highest and the lowest probability of samples inside a hyper-rectangle is bound by the following condition:
\begin{equation}
 \frac{f_\mathrm{max}}{f_\mathrm{min}} \leq t \; .
\label{eq:threshold}
\end{equation}
Although this threshold is user-selectable, by default it is equal to $500$ (an example of integration with different threshold values is shown in Sec.~\ref{subsection:multivariate_normal}). 


To create $M$ regions for integration, we select the seed point $\lambda^{\textrm{seed}}_i$ with the overall largest value $f(\lambda^{\textrm{seed}}_i)$ and follow the steps below. Then we recursively repeat the same procedure $M-1$ times using the remaining seed points.
\begin{enumerate}
    \item The algorithm starts by building a small hyper-cube $\tilde{\Delta}_i$ around the selected seed point $\lambda^{\textrm{seed}}_i$, see Fig.~\ref{fig:seed}.
    \item This hyper-cube is then incrementally either increased or decreased in size, until the probability ratio of contained samples matches the threshold $t$ within some tolerance, or until it contains more than one percent of the total samples.
    \item The faces of the hyper-cube are then iteratively adjusted (expand or contract), to adapt to the density of the contained samples, while enforcing the condition $f_{max}/f_{min} \leq t$. This step turns the $D$-dimensional hyper-cube into a $D$-dimensional hyper-rectangle.
    This hyper-rectangle adaptation algorithm continues as long as changes to the hyper-rectangle's faces are accepted. The stopping criterion is based on the fraction of samples accepted or rejected compared to expectation from the volume change. However, the hyper-rectangle adaption algorithm always ensures that no modification to the hyper-rectangle's faces are made if such a modification would result in $f_{\mathrm{max}}/f_{\mathrm{min}} > t$.
\end{enumerate}
Figure~\ref{fig:visualizationhmi} shows the resulting set of $M$ hyper-rectangles for our example.

A more technical description of the algorithm can be found in \ref{sec:volume_creation}.

\subsection{Integral Estimates}

Once $M$ integration regions are defined, we can compute the integral estimates $\hat{I}^A_i$ for each $\Delta^B_i$, according to Eq.~\ref{eq:ourHME2}. The procedure is the same for $\hat{I}^B_i$, so we shall drop the superscripts $A$ and $B$ in the following two sub-sections. The two resulting, separate estimates $\hat{I}^A$ and $\hat{I}^B$ will then be combined in Sec.~\ref{subsec:final_estimate} to obtain the final estimate.

From the distribution of all estimates $\hat{I}_i$ we select only the 68\% central percentile to reject outliers---a procedure that was empirically found to work well. This is indicated in Fig.~\ref{fig:visualizationhmi} labeled as ``accepted" and ``rejected" rectangles.
We proceed to combine the remaining estimates $\hat{I}_i$ into a single estimate $\hat{I}$ using a robust and unbiased estimator for the combination of correlated measurements as suggested in \cite{ref:schmelling}.


\begin{equation}
\hat{I} = \sum_{i} w_i \hat{I}_i \;\;\;\;\;\;\;\;\;\;\;\;\;\;\;\;
\sigma^2(\hat{I}) = \sum_{i,j} w_i w_j \bar{\sigma}_{ij},
\end{equation}
where the weights $w_i$ are defined as:
\begin{equation}
w_i = \frac{\frac{1}{\bar{\sigma}_i^2}}{\sum_{j}\frac{1}{\bar{\sigma}_j^2}}.
\end{equation}
The variances $\bar{\sigma}_i^2 \equiv \bar{\sigma}_{ii}$ and covariances $\bar{\sigma}_{ij}$ of the mean assigned to integration regions $\Delta_i$ and $\Delta_j$ are estimated in the next section.



\subsection{Covariance Estimate}
The following procedure is used to estimate the covariance between individual integral estimates $\hat{I}_i$:
\begin{enumerate}
\item We partition $\{\Lambda\}$ into a number \textit{S} of subsets $\{\Lambda_1\}$, $\{\Lambda_2\}$ ... $\{\Lambda_S\}$ chosen in a way that reduces their correlation. The default value for $S$ is $10$.
\item  Separate estimates $\hat{I}_{i,k}$ ($k$ indexes the $S$ partitions) of the integral are then performed for all sample subsets $\{\Lambda_k\}$ resulting in $S$ integral estimates for each subspace $\Delta_i$:
$$\begin{bmatrix} \hat{I}_{i,1} &  \hat{I}_{i,2} &  ... &  \hat{I}_{i,S} \end{bmatrix}$$
\item The resulting sample variance can then be written as 
$$\sigma^2_i = \frac{1}{S-1}\sum_{k=1}^{S} \left(\hat{I}_{i,k} - \mu_i \right)^2.$$
with 
$$\mu_i=\frac{1}{S} \sum_{k=1}^{S} \hat{I}_{i,k}$$ 
\item The estimate of the variance of $\hat{I}_i$ is then
$\bar{\sigma}_i^2 = \frac{\sigma_i^2}{S}$.
\item We found that following a similar procedure to estimate the covariance ($\sigma_{ij} = \frac{1}{S-1}\sum_{k=1}^{S} (\hat{I}_{i,k} - \mu_i)(\hat{I}_{j,k} - \mu_j)$) misrepresented the correlation between the samples in different hyper-rectangles. We suspect this being the case due to an underestimate of the overlap of the hyper-rectangles.

We therefore use a different and more robust estimate of the covariances given by:
$$ \bar{\sigma}_{ij} = \rho_{ij}\bar{\sigma}_i\bar{\sigma}_j, $$
where 
\[\rho_{ij}\ = \frac{W_\cap(i,j)}{W_\cup(i,j)}\]
with $W_\cap(i,j)$ and $W_\cup(i,j)$ the number of weighted samples in $\Delta_i\cap\Delta_j$ and $\Delta_i\cup\Delta_j$ respectively.
\end{enumerate}

\subsection{Final AHMI Integral Estimate}\label{subsec:final_estimate}

As we have now obtained two values of $\hat{I}$ and two variances $\sigma^2(\hat{I})$ from the two sets $\{\Lambda^A\}$ and $\{\Lambda^B\}$, we combine those into the final result like
\[
\hat{I} = \frac{\hat{I}^A/\sigma_A^2+\hat{I}^B/\sigma_B^2}{1/\sigma_A^2+1/\sigma_B^2}
\]
with variance estimate

\[\sigma^2 = \left(\frac{1}{\sigma_A^2} + \frac{1}{\sigma_B^2}\right)^{-1}.\]


\section{Benchmark Examples}
\label{sec:benchmark}
To validate our algorithm, we apply it to estimate the integral of several test functions in varying dimensionality (up to $d=25$) for which an analytic (or accurate numerical) solution of the integral value is available. 

The test functions were chosen to pose different challenges to the algorithm. For our first test problem, we start with the canonical example of a multivariate normal distribution. For the second test case we look at Gaussian shells, for which the mode of the distribution does not lie on a single point but has infinite modes on $d-1$-dimensional surface. Next we study the heavy-tailed Cauchy distribution with multiple modes, and in the end explore the asymmetric ``Funnel'' function.
Additional information of the number of integration volumes used and the computation time are only provided for the first example, as these are very similar for the other three examples.

The samples $\{\Lambda\}$ on which our integration is based are obtained from Metropolis-Hastings MCMC~\cite{ref:github_bat}, and in the case of the multivariate normal from \textit{i.i.d.} sampling. The sample size is fixed to $2\cdot10^6$ for Gaussian shells example and it is equal to $10^6$ for other examples. 



\subsection{Multivariate Normal Distribution}
\label{subsection:multivariate_normal}

\begin{figure}[b!]
    \centering
    \includegraphics[width=\textwidth]{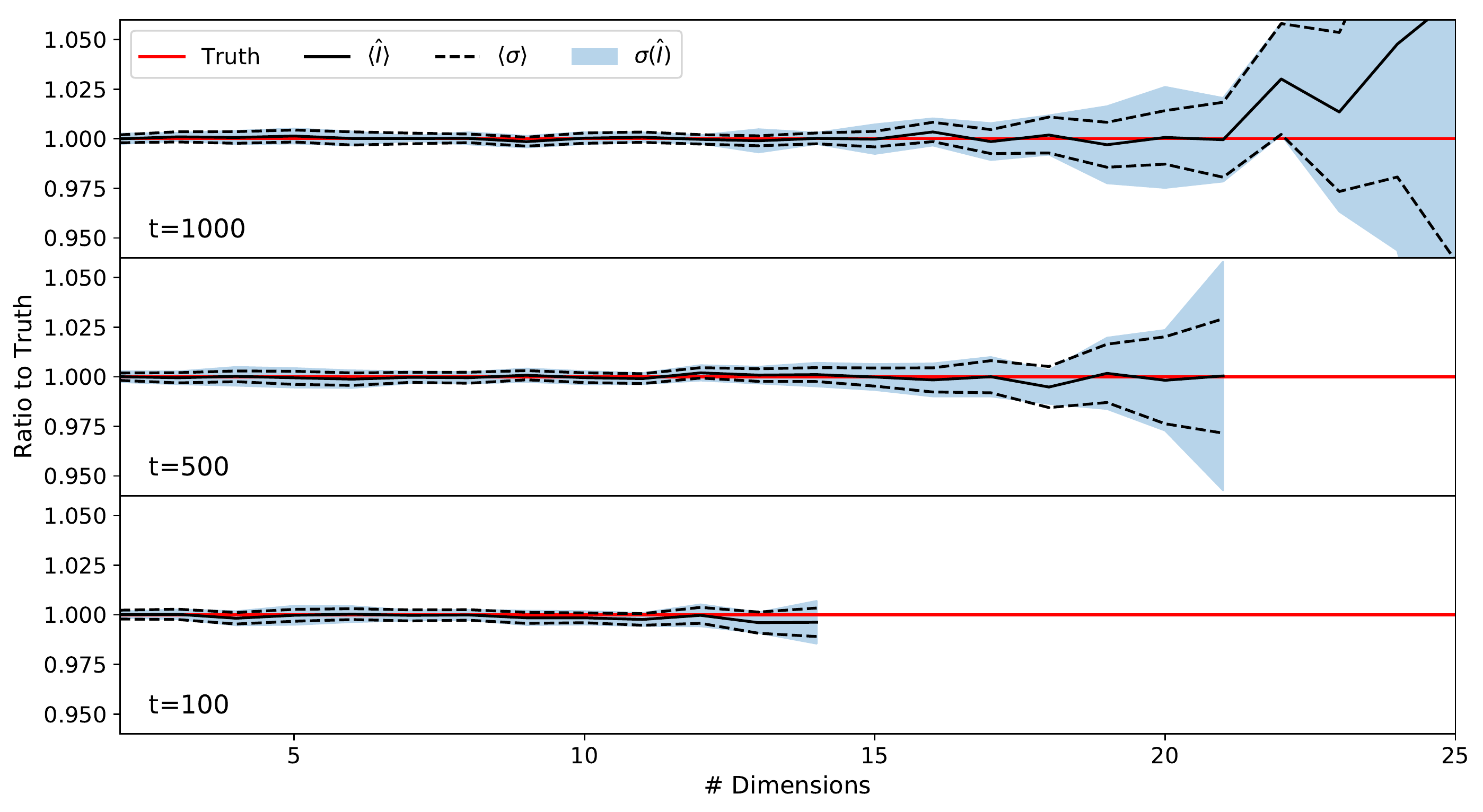}
    \caption{AHMI value for the multivariate normal distribution as ratio to the true integral shown as a function of dimensionality for three threshold values $t=[100, 500, 1000]$. The solid black lines give the mean result over ten independent trials and the shaded bands show the standard deviation of these trials. The dashed lines show the average errors reported by AHMI. Samples are obtained from \textit{i.i.d.} sampling, $10^6$ for each run.}
    \label{fig:normal_benchmark}
\end{figure}

\begin{figure}
    \centering
    \includegraphics[width=0.9\textwidth]{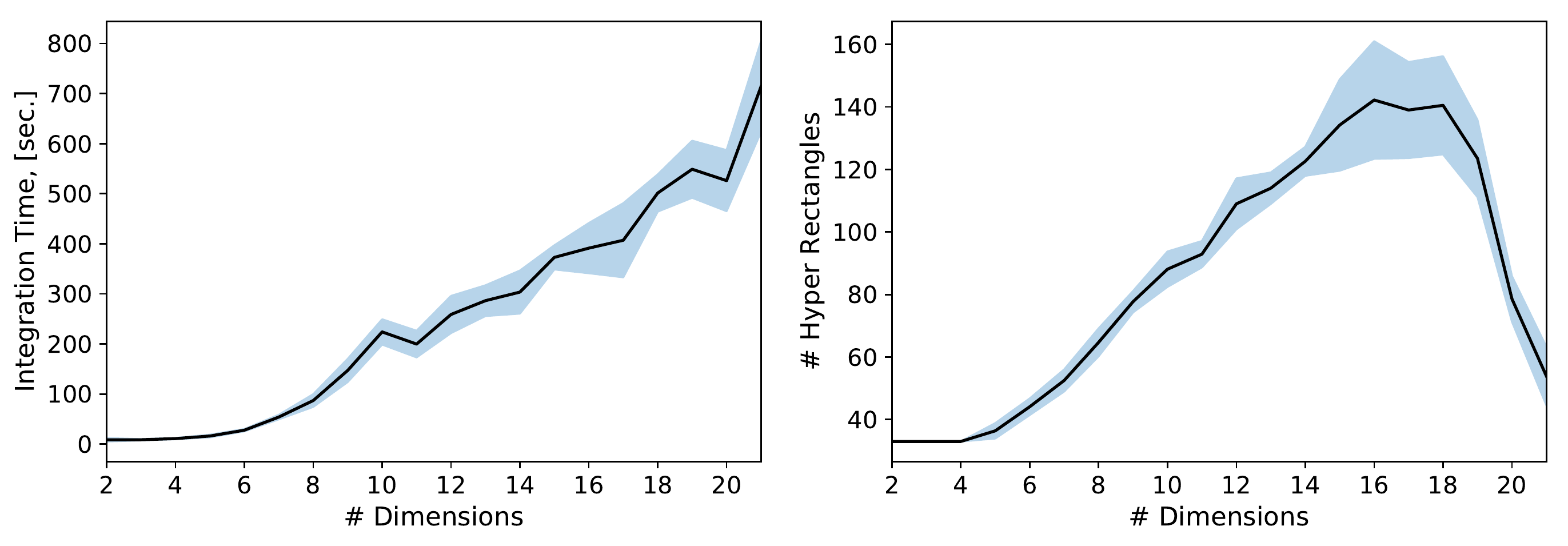}
    \caption{Left: Average AHMI execution time for the multivariate normal distribution ($t=500$) in total CPU seconds (run on a system with a 2.3 GHz Intel Xeon 6140 processor) as a function of dimensionality.
     Right: Average number of hyper-rectangles used by AHMI for computing the integral estimate for the multivariate normal distribution as a function of dimensionality.}
    \label{fig:time_cubes}
\end{figure}

The first test case is a unit normal distribution (centered at zero, width one) in two up to 25 dimensions. The samples input to AHMI are obtained from \textit{i.i.d.} sampling. The resulting integral estimates are shown in Fig.~\ref{fig:normal_benchmark} as a function of the dimensionality for three different threshold values $t=[100, 500, 1000]$. 

For both threshold values, $t=[500, 1000]$, we get unbiased and consistent results up to around $21$ dimensions, after which results start to become positively biased for $t=1000$. This bias seems to be intrinsic to the method itself or our implementation of the algorithm and subject of future studies. For the threshold value $t=100$ we get an unbiased integral estimate up to 14 dimensions, and after that hyper-rectangles can no longer be created. The default value of $t=500$ was used for the remaining examples below.

In Fig.~\ref{fig:time_cubes} we show the execution time of the algorithm, and the number of hyper-rectangles used for integration for the threshold value $t=500$. The execution time rises with the number of dimensions almost linearly. The change in slope at low dimensionality is likely due to CPU caching behaviour. The number of hyper-rectangles starts to decay after $18$ dimensions indicating that there exist fewer hyper-rectangles that satisfy Eq.~\ref{eq:threshold}. 


\begin{figure}[b!]
    \centering
    \includegraphics[width=0.65\textwidth]{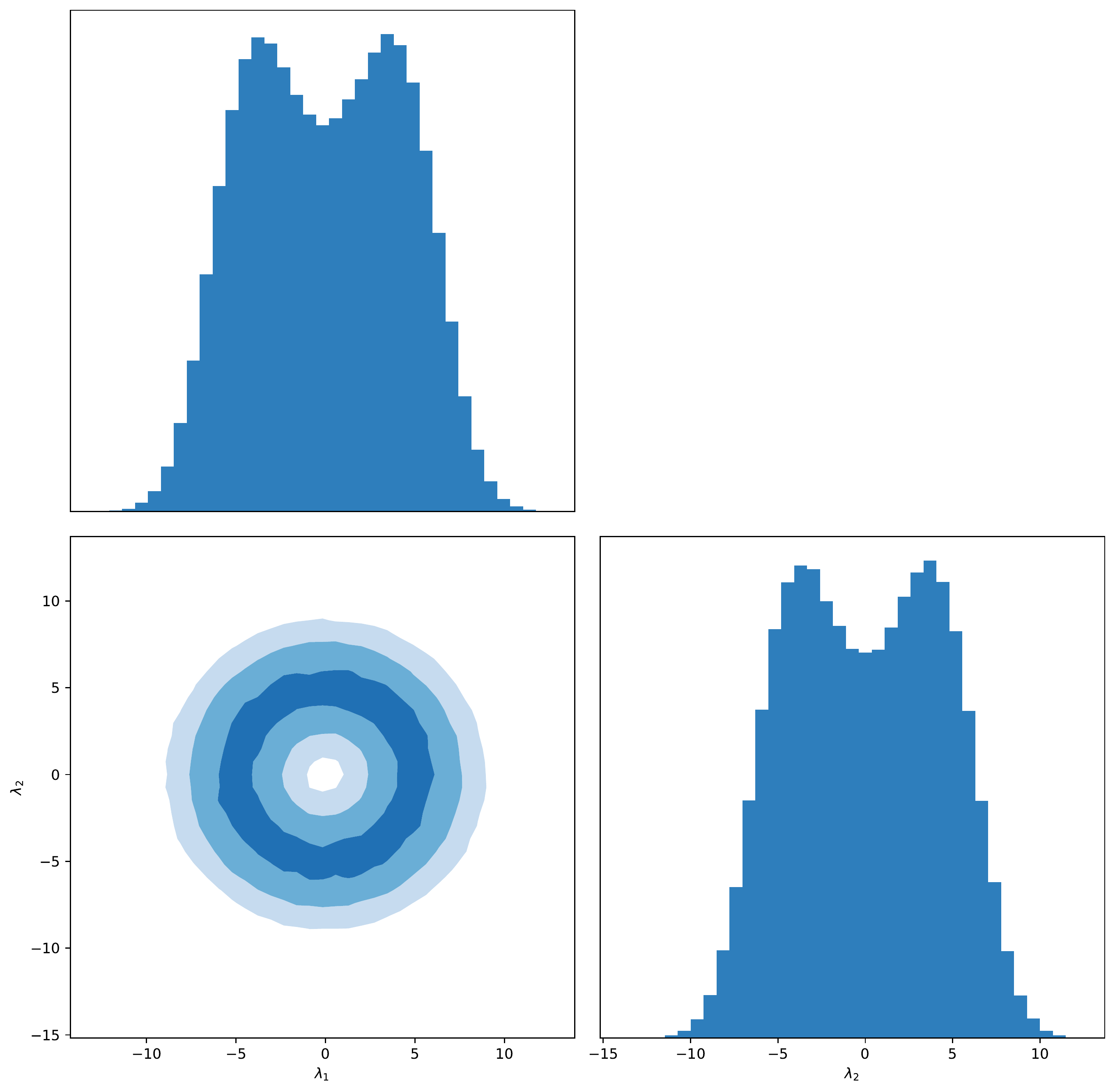}
    \caption{One and two dimensional distributions of samples along the first two dimensions of the Gaussian shell target function.}
    \label{fig:shell_dist}
\end{figure}

\subsection{Gaussian Shell Distribution}

\begin{figure}
    \centering
    \includegraphics[width=\textwidth]{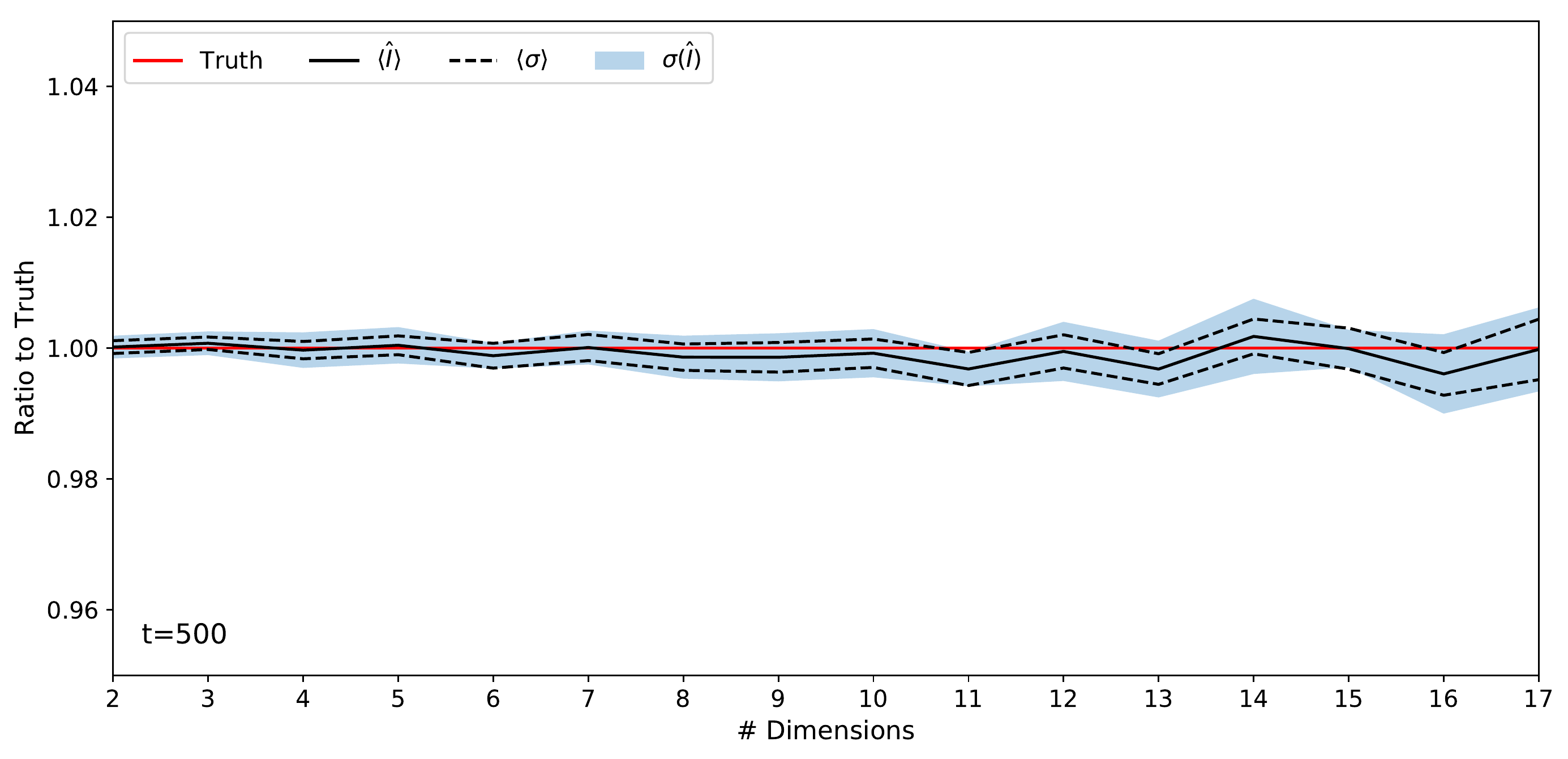}
    \caption{AHMI value for the Gaussian shell distribution as ratio to the true integral shown as a function of dimensionality. The solid line gives the mean result over ten independent trials and the shaded band show the standard deviation of these trials. The dashed lines show the average errors reported by AHMI. The samples are obtained from  Metropolis-Hastings MCMC and the sample size is $2\cdot10^6$.}
    \label{fig:shell_benchmark}
\end{figure}

The functional form 
was given in Eq.~\ref{2dshell} and an example distribution in the first two dimensions is shown in Fig.~\ref{fig:shell_dist}. The AHMI algorithm results (Fig. ~\ref{fig:shell_benchmark}) shows a similar behaviour as for the multivariate normal distribution for this more complicated test function than . However, integration was possible up to 17 dimensions. Up to that point the integral estimates, including errors, are well behaved.

\subsection{Multimodal Cauchy Distribution}

\begin{figure}
    \centering
    \includegraphics[width=0.8\textwidth]{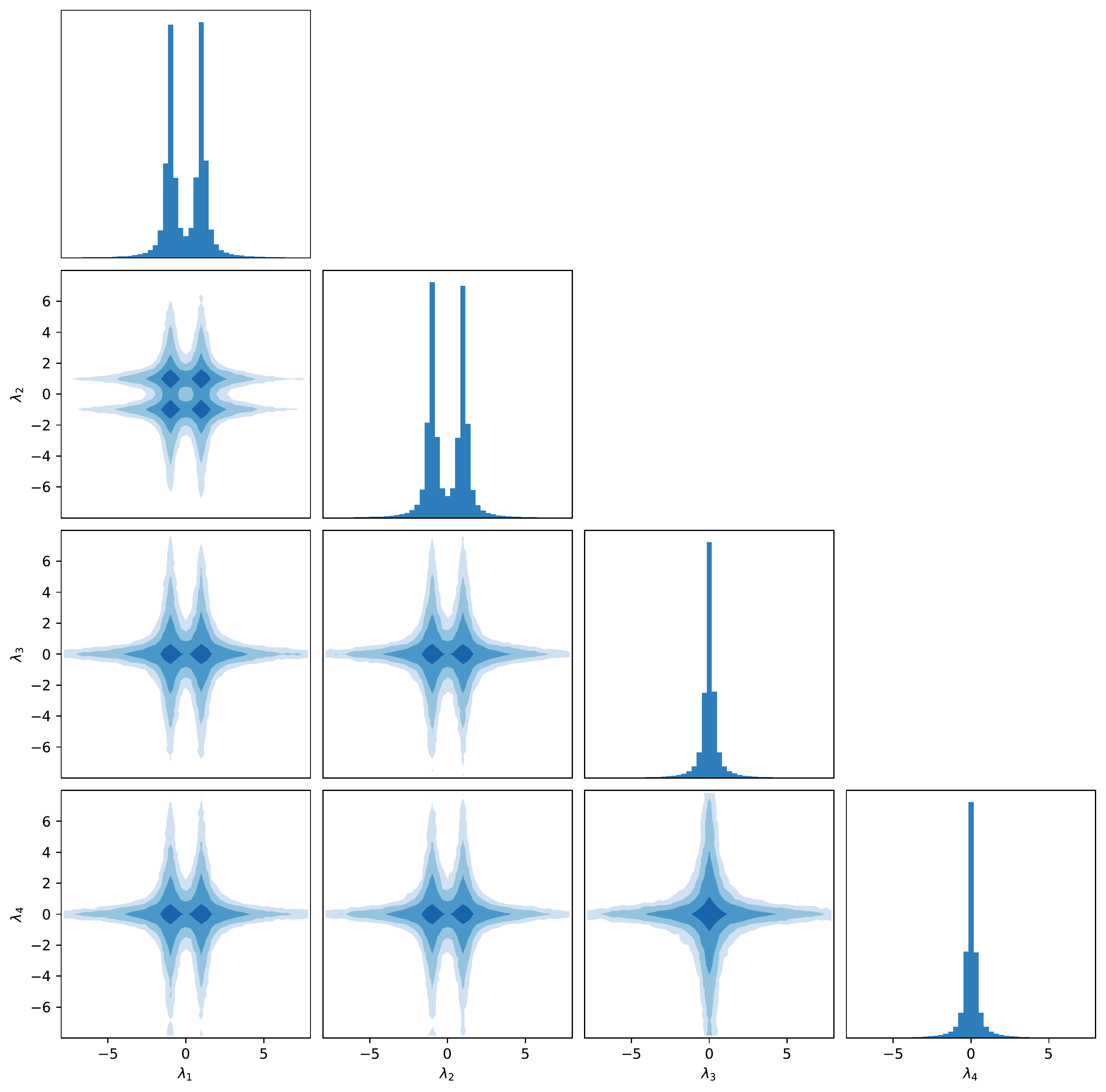}
    \caption{One and two dimensional distributions of samples along the first four dimensions of the multimodal Cauchy target function.}
    \label{fig:cauchy_dist}
\end{figure}

\begin{figure}
    \centering
    \includegraphics[width=0.9\textwidth]{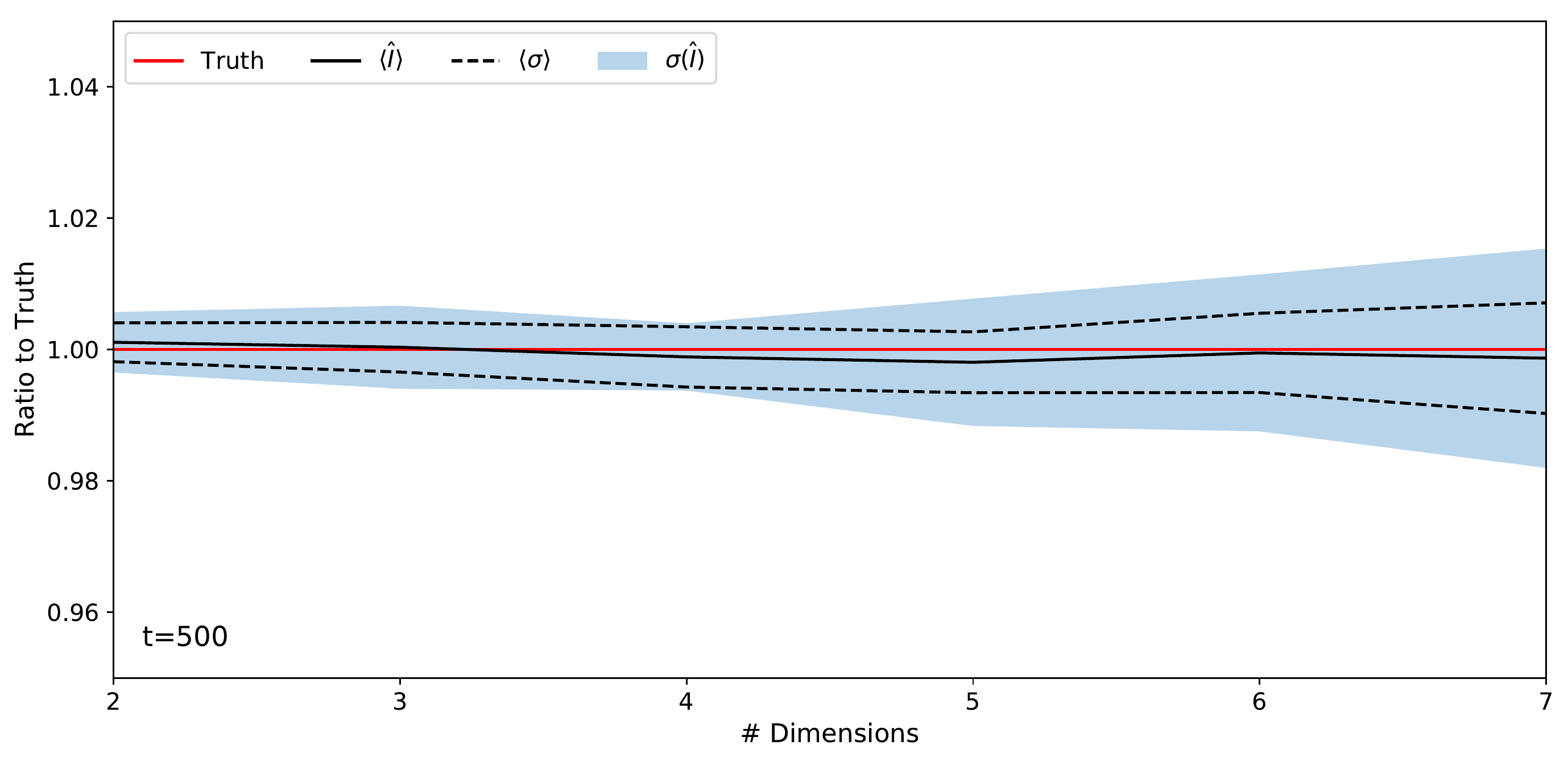}
    \caption{AHMI value for the multimodal cauchy distribution as ratio to the true integral shown as a function of dimensionality. The solid line gives the mean result over twenty independent trials and the shaded band show the standard deviation of these trials. The dashed lines show the average errors reported by AHMI.}
    \label{fig:caushy_benchmark}
\end{figure}

The Cauchy distribution, with its heavy tails, is a notoriously difficult problem and used here to point out possible weaknesses of our algorithm. We further increase complexity by using four separate, shifted Cauchy distributions creating multiple modes. The functional form can be written as
\begin{equation}
    f\left ( \lambda \right )  = \prod_{i=1}^2 \frac{1}{2} \left [ \text{Cauchy}\left ( \lambda_i \mid  \mu, \sigma \right ) + \text{Cauchy}\left ( \lambda_i \mid - \mu, \sigma \right ) \right ] \cdot \prod_{j=3}^n \text{Cauchy}\left ( \lambda_j \mid  0, \sigma \right ),
\end{equation}
where $\mu=1, \sigma=0.2$ and $n$ is a dimensionality of $\lambda$. 
An example of this traget distributions is provided in Fig.~\ref{fig:cauchy_dist}. The integration region extends from $[-8, 8]$ in each dimension. Our results are collected in Fig.~\ref{fig:caushy_benchmark} and indicate that integration is possible up to seven dimensions, given the fixed sample size of $10^6$. For this range, the AHMI results are very reliable.

\subsection{Funnel Distribution}

\begin{figure}
    \centering
    \includegraphics[width=0.7\textwidth]{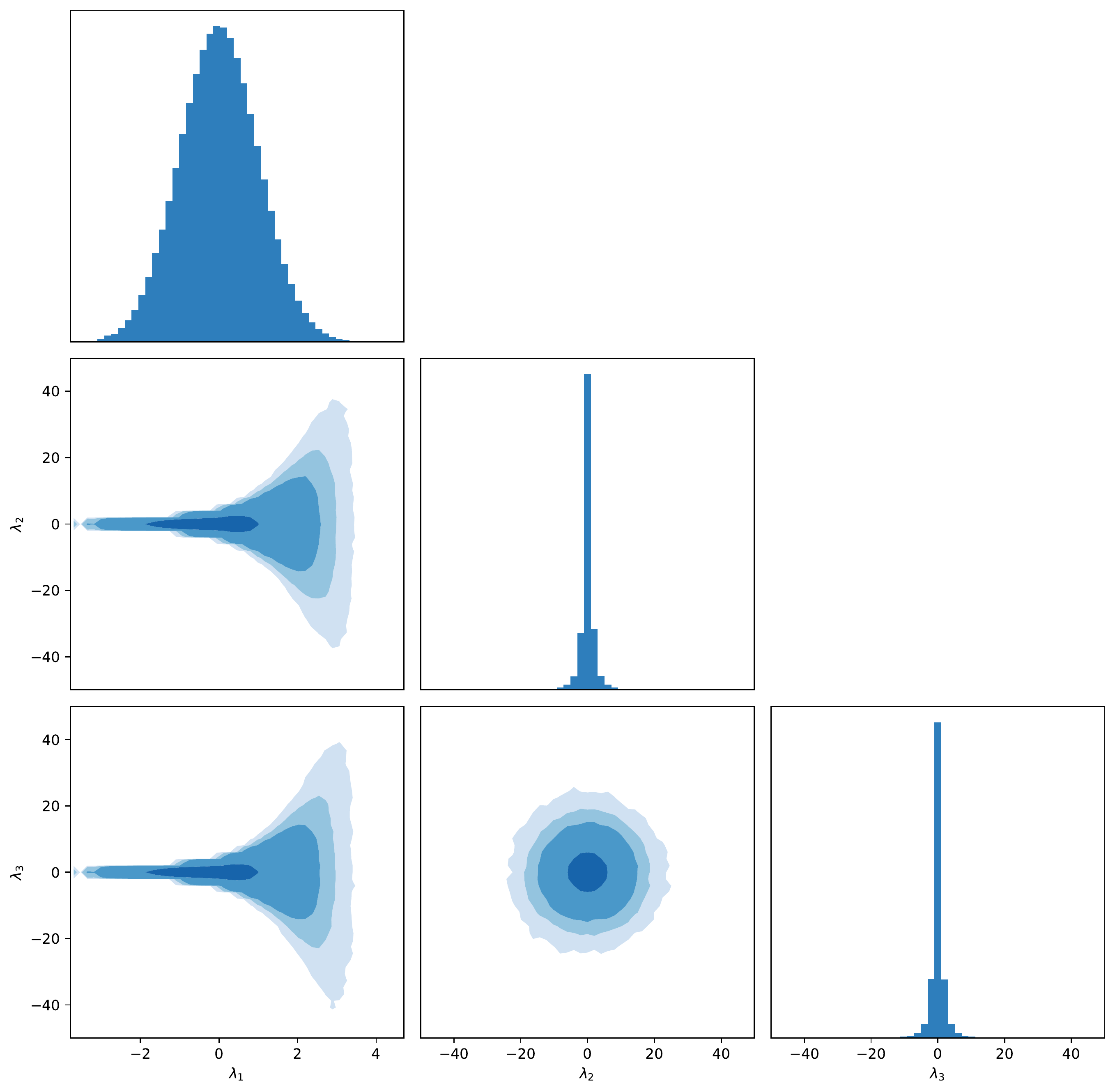}
    \caption{One and two dimensional distributions of samples along the first three dimensions of the Funnel target function.}
    \label{fig:funnel_dist}
\end{figure}

\begin{figure}[t!]
    \centering
    \includegraphics[width=0.9\textwidth]{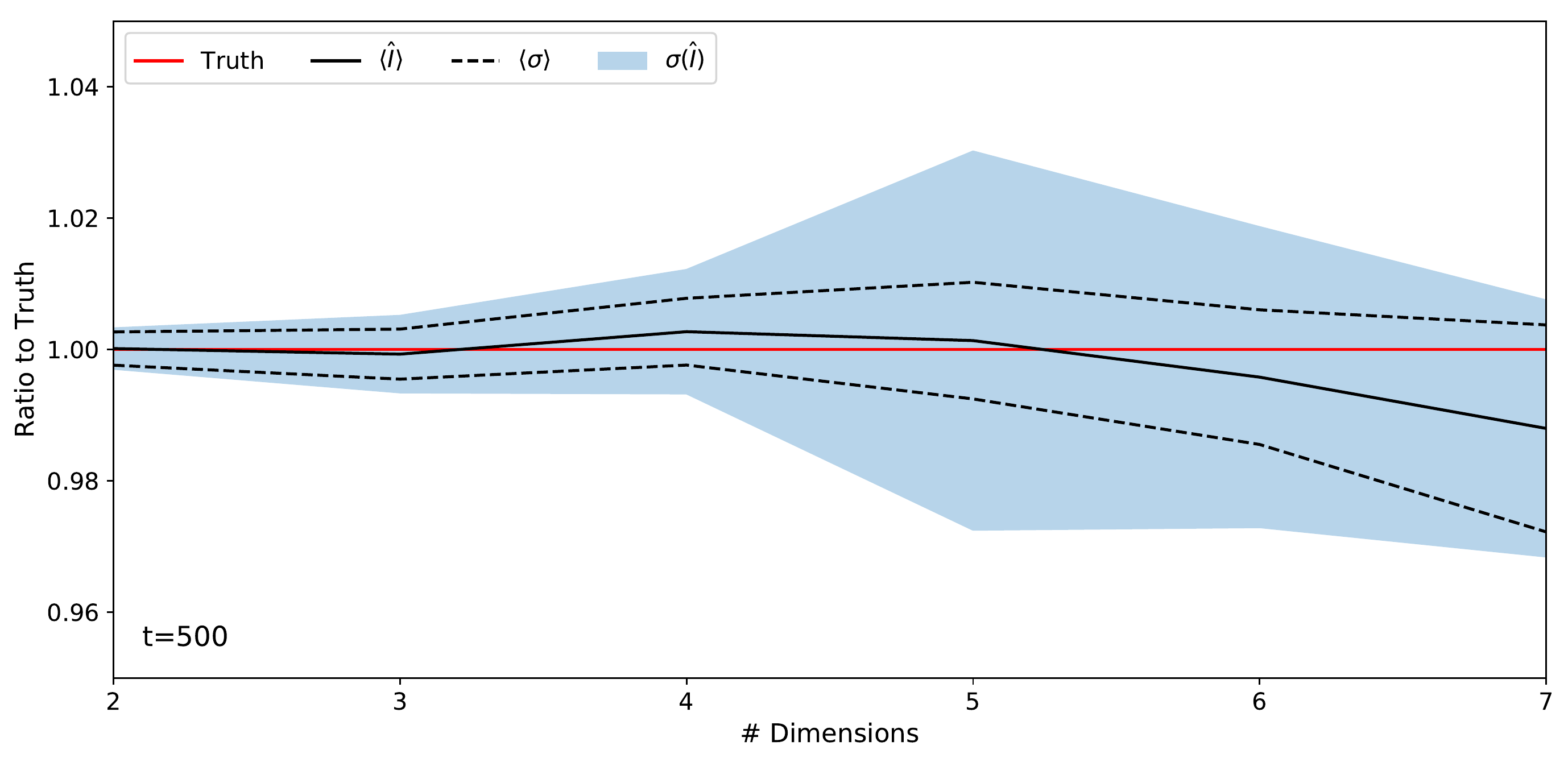}
    \caption{AHMI value for the "Funnel" distribution as ratio to the true integral shown as a function of dimensionality. The solid line gives the mean result over twenty independent trials, with the shaded band showing the standard deviation of these trials. The dashed lines show the average errors reported by AHMI.}
    \label{fig:funnel_benchmark}
\end{figure}

The final problem we study is the so-called ``Funnel" distribution, that is described in \cite{Jia:2019wsj}. The functional form of this distribution can be written as 
\begin{equation}
    f\left (\lambda \right )  =  \mathcal{N}\left ( \lambda_1 \mid 0, a^2 \right ) \prod_{i=2}^n  \mathcal{N}\left ( \lambda_i \mid 0, \exp\left ( 2b\lambda_1 \right ) \right ),
\end{equation}
where $a=1$, $b=0.5$ and $n$ is a dimensionality of $\lambda$. An example of the distribution in its first three dimensions on the parameter range $[-50, 50]$ is provided in Fig.~\ref{fig:funnel_dist}.
The results (Fig.~\ref{fig:funnel_benchmark}) show a similar performance as for the previous example with reliable estimates up to seven dimensions.

\section{Conclusion}
We have developed an Adaptive Harmonic Mean Integration (AHMI) algorithm that can be used to integrate a non-normalized density function using the samples drawn according to the 
probability distribution proportional to the function.  The fundamental assumption is that the sampling algorithm has faithfully produced samples from this distribution.  Given this, the AHMI algorithm can be used to produce both an  estimate of the integral of the function over its full support as well as an estimate of the uncertainty of the integral. In this first implementation of the AHMI algorithm, finite hyper-rectangles are generated in the whitened space of the samples covering the full support.  The adaptive algorithm ensures that the range of function values enclosed by the hyper-rectangles is limited such that the variance of the integral results are moderate.  This allows for reliable results both for the integral values as well as for reliable uncertainty estimates.

The algorithm has been tested on a number of examples and found to perform well in all cases.   The use of hyper-rectangles however limits the applicability to a not-too-large number of dimensions ($\approx 20$ in the case of the multivariate normal distribution).  An extension of the technique to differently shaped integration regions is possible.





\section*{Acknowledgements}
This research was supported by the European Union’s Framework Programme for Research and Innovation Horizon 2020 (2014-2020) under the Marie Sklodowska-Curie Grant Agreement No.765710 and the Deutsche Forschungsgemeinschaft (DFG, German Research Foundation) under Germany´s Excellence Strategy – EXC-2094 – 390783311.

\appendix

\section{Hyper-rectangle Generation Algorithm}
\label{sec:volume_creation}

The hyper-rectangle creation process, described in detail in this section, tries to find hyper-rectangle shaped integration volumes, that adapt to the $d$-dimensional data as best as possible. As starting point, this algorithm uses the previously determined starting samples. The hyper-rectangle creation process starts by building a $d$-dimensional hypercube around one of the starting samples. Thereafter, the faces of the hyper-cube are iteratively adjusted, thus turning the $d$-dimensional hyper-cube into a $d$-dimensional hyper-rectangle, sometimes also called $d$-orthotope.

\subsection{Initial Hyper-Cube Creation}
An initial hyper-cube with edge length 1 is centered at one of the starting samples. If this initial hyper-cube contains more than 1\% of the total number of samples, then the hyper-cube's edge length gets multiplied by a factor of $0.5^{1/d}$, thus halving the hypercube's volume. This volume-halving process continues until the hyper-cube contains less than 1\% percent of the total number of samples.
Thereafter, the hyper-cube is incrementally either increased or decreased, until its probability ratio matches the probability ratio threshold $t$ within a given tolerance or until it contains more than one percent of the total samples (again). This initial hyper-cube is the basis for further modifications of its individual faces, which then turn the hyper-cube into a hyper-rectangle.

\subsection{Hyper-Rectangle Adaptation}
Subsequently, the faces of the hyper-rectangle, created as described above, get adjusted
in a way that the hyper-rectangle adapts to the samples, representing the target density
function, as best as possible. The algorithm, that modifies the hyper-rectangle's faces,
iteratively adjusts the lower and the upper face along one dimension at a time. The algorithm keeps iterating through all dimensions of the
parameter space. During each iteration, both faces of the hyper-rectangle in the current
dimension are adjusted individually. Therefore, the algorithm may either increase or
decrease the hyper-rectangle's volume by offsetting the hyper-rectangle's face. The offset
is chosen in a way, that the volume is changed by 10\% (configurable). The algorithm
expects the number of samples within the hyper-rectangle to increase proportionally with
a volume increase. Therefore, the algorithm is able to perform larger volume changes than
10\% to ensure that an addition of twenty new points may be expected by this change.
Such a volume increase is performed only, if a sufficient number of samples is added. This
is the case, when the volume change includes a new sub-volume of high probability mass.
Correspondingly, a volume decrease is performed only, if very few or no samples are lost,
which happens when a sub-volume of low probability mass is excluded. Such sub-volumes
of low probability mass are undesirable to be within the hyper-rectangle's volume and
should be avoided, because they may falsify the integration estimate.
To determine whether the number of points added or removed are sufficient to perform a
volume change, a parameter $\mu$ is employed. This parameter $\mu$ describes the expectation
on the number of samples, that are to be added or removed by a volume change. $\mu$ is
data-dependent and gets calculated before the hyper-rectangle creation process. 
Details on how this parameter is obtained from the data are explained in \ref{sec:adaption}.
The threshold, for performing a volume change (increase or decrease) to the hyper-rectangle's faces is given by the following formulas:
\begin{align}
\label{eq:thre  shold}
p_{change,\ increase} &> 1 + \frac{V_{change}}{\mu} \\
p_{change,\ decrease} &> 1 - \frac{V_{change}}{\mu}
\end{align}
where $p$ represents for the percentual increase in the number of samples inside the hyper-rectangle after a volume increase or decrease. The parameter $V$ represents the percentual volume change.
\begin{align}
V_{change} &= \frac{V_{new}}{V_{old}} \\
p_{change} &= \frac{p_{new}}{p_{old}}
\end{align}

However, the hyper-rectangle adaption algorithm always ensures, that no modification to the hyper-rectangle's faces are made, that would increase the probability ratio beyond the probability ratio threshold $t$, required to preserve numerical stability.

\subsection{Threshold for Adapting the Hyper-Rectangle}
\label{sec:adaption}
The parameter $\mu$ describes the expectation on the number of samples, that are added or removed by a volume change and is employed in the threshold, which determines whether a proposed volume increase or decrease is performed (see Eq.~\ref{eq:threshold}). In order to obtain this threshold, the first step is to generate four test hyper-cubes around the global mode with different volumes, which are chosen in a way that they contain multiples (onefold, two-fold, four-fold and eight-fold) of a certain number of samples $n$, where $n$ is set to four times the number of dimensions. This procedure is then repeated for the ten samples with highest probability as well.
Thereafter, for each of the ten samples with highest probability (including the sample at the global mode), three ratios are calculated, which represent the percentual volume change divided by the percentual number of sample change for hyper-cubes containing a different number of samples.
\begin{equation}
    \mu_i=\frac{V_i}{p_i}
\end{equation}
Thus the hyper-cube containing the eigth-fold number of samples $n$ is compared to the hyper-cube containing the four-fold number of samples $n$ (yielding ratio $\mu_1 = V_1/p_1$). Next, the
hyper-cube containing the four-fold number of samples $n$ is compared to the hyper-cube containing the two-fold number of samples $n$ (yielding ratio $\mu_2$) and so on. Thus, each of the chosen samples has three ratios assigned, each of them describing how the number of samples changes, when the hyper-cube's volume changes. The final parameter $\mu$ is then based on the mean value of these ratios. The parameter $\mu$ gets scaled up by $k = 4$ to allow for more tolerant volume increases or decreases because the actual hyper-rectangles' volumes are much larger than the test hyper-cubes' volumes which mostly are located near the global mode, which generally is very densely populated with samples.

\begin{equation}
    \mu = \left( \frac{1}{n} \sum_{i=1}^n{\mu_i-1}\right) \cdot k +1
\end{equation}

\bibliographystyle{ba}

\end{document}